\documentstyle[12pt, epsf]{article}
\begin{document}
\textwidth 160mm
\textheight 240mm
\topmargin -20mm
\oddsidemargin 0pt
\evensidemargin 0pt
\newcommand{\beq}{\begin{equation}}
\newcommand{\eeq}{\end{equation}}
\newcommand{\be}{\begin{equation}}
\newcommand{\ee}{\end{equation}}
\newcommand{\lll}{\lambda}
\def\Journal#1#2#3#4{{#1} {\bf #2}, #3 (#4)}

\def\NCA{\em Nuovo Cimento}
\def\NIM{\em Nucl. Instrum. Methods}
\def\NIMA{{\em Nucl. Instrum. Methods} A}
\def\NPB{{\em Nucl. Phys.} B}
\def\PLB{{\em Phys. Lett.}  B}
\def\PRL{\em Phys. Rev. Lett.}
\def\PRD{{\em Phys. Rev.} D}
\def\ZPC{{\em Z. Phys.} C}
\def\st{\scriptstyle}
\def\sst{\scriptscriptstyle}
\def\mco{\multicolumn}
\def\epp{\epsilon^{\prime}}
\def\vep{\varepsilon}
\def\ra{\rightarrow}
\def\ppg{\pi^+\pi^-\gamma}
\def\vp{{\bf p}}
\def\ko{K^0}
\def\kb{\bar{K^0}}
\def\al{\alpha}
\def\ab{\bar{\alpha}}
\def\be{\begin{equation}}
\def\ee{\end{equation}}
\def\bea{\begin{eqnarray}}
\def\eea{\end{eqnarray}}
\def\CPbar{\hbox{{\rm CP}\hskip-1.80em{/}}}


\begin{flushright}
IHES/P/01/47  \\
ITEP-TH-58/01\\
LPTHE-01-56  \\
LPT-Orsey-01/103 \\
OUTP-01-56P \\
hep-th/0111013
\end{flushright}

\setcounter{footnote}{0}

\setcounter{equation}{0}
\centerline{
\bf Dualities in  Quantum Hall System  }
\centerline{
\bf and  Noncommutative Chern-Simons  Theory}
\vspace{0.5cm}
\centerline{ A. Gorsky$^{\,a,b}$, I.I. Kogan$^{\,a,c,d,e}$ and 
C. Korthels-Altes$^{\,f}$ }
\begin{center}
$^a$ {\em Institute of Theoretical and Experimental Physics, \\
B.Cheremushkinskaya 25, Moscow,  117259, Russia }\\
$^b$ {\em LPTHE, Universite' Paris VI, \\
4 Place Jussieu, Paris, France}\\
$^c$
{\em  Theoretical Physics, Department of Physics, Oxford University\\
1 Keble Road, Oxford, OX1 3NP, UK } \\
$^d$ 
{\em IHES, 35 route de Chartres \\
91440, Bures-sur-Yvette,  France }\\
$^e$
{Laboratoire de Physique Th\'eorique%
Universit\'e de Paris XI, \\
91405 Orsay C\'edex, France
}\\
$^f$
{\em Centre Physique Theorique, \\
au CNRS, Case 907, Luminy 13288 Marseille, France}
\end{center}
\begin{abstract}
We discuss different dualities  of QHE in the
framework of the noncommutative Chern-Simons theory.
First, we consider the Morita or T-duality transformation on the torus  
which maps the abelian noncommutative CS description of QHE
on the torus into the nonabelian commutative description on the dual 
torus.  It is argued that
the Ruijsenaars integrable many-body system provides the
description of the QHE with  finite amount of electrons
on the torus. The new IIB brane picture 
for the QHE is suggested and 
applied to  Jain and generalized hierarchies. This picture naturally
 links 2d $\sigma$-model and 3d CS description of the QHE.   All
duality transformations  are identified  in the brane setup and can 
be related with the mirror symmetry and S duality. We suggest a brane 
 interpretation of the plateu transition in IQHE in  which a critical
point is  naturally described by  $SL(2,R)$ WZW model.
\end{abstract}

\section{Introduction}
Non-commutative field theories (see \cite{dn} for a review and references 
therein)are with us since some time and 
they owe their raison d'\^etre to a variety of reasons.
The most recent one is string theory \cite{sw}
 which in a particular region of parameter 
space  produces non-commutative field theory in very much the same 
way as the lowest Landau level starts 
to produce non-commuting coordinates in a 
very strong magnetic field. This phenomenon was studied in the thirties by 
Moyal, Peierls and many others in the setting of an additional periodic
potential.

Another way of producing non commutative gauge theory occurred in the
phenomenon of transmutation of internal (color) to external (space time) degrees 
of freedom. This was popular in the early eighties and went by the
name of twisted Eguchi-Kawai models~\cite{okawa}. The idea was to have a sequence of conventional
SU(N) gauge theories with twisted boundary conditions on a finite lattice 
(or even one point lattice) and to produce a NC theory on a large lattice.

The interaction between the photons in the NC theory vanishes linearly
in the momenta, when they are small with respect to the NC scale $\mu$
in contrast to the ordinary YM coupling.
On the other hand one can express the free energy of the NC theory
in a periodic box of size $L$.L is supposed to be larger, by an integer
factor N, with respect to the NC scale. Then, to all orders 
in perturbation theory, the free energy of NCU(1) equals the free energy of conventional 
SU(N) gauge theory in a twisted box. The size of this 
box is small with respect to the NC scale. And the coupling of the NC theory
rescales by this same factor $N$. This is the simplest form of Morita
duality~\cite{morita}.

Another  raison d'\^etre of NCU(1) theory was its connection with large N theory.
In fact Gonzalez-Arroyo et al~\cite{gonzalez} showed for fixed cut-off that the non-planar sector is suppressed in the limit of
infinitely large NC length scale $\mu^{-1}$, whereas the planar sector
is identical. Only for exceptional momenta the non-planar sector can
survive.

When the parameter of noncommutativity 
is rational one can map the noncommutative
gauge theories to the commutative gauge theories with  twisted
boundary conditions.  Under the Morita duality transformation 
in d dimensions the topological
numbers of the bundles, namely the rank, flux and the Pontryagin number
are transformed under SO(d,d,Z) action. The
noncommutativity parameter $\theta$ gets mapped into the twist
in the dual commutative YM theory. There exists an explicit mapping
of the gauge fields under this transformations \cite{mak}.

It was also recognized recently that noncommutative Chern-Simons theory
provides the appropriate language for the description
of the FQHE \cite{susskind}.
The particle density and the filling factor get mapped into the
noncommutativity parameter  and the coefficient in front of the CS action.
Generalization of the picture for a finite amount of electrons
having the finite extent on the plane was found in \cite{poly}.
It appears that the corresponding regularized model is equivalent
to the rational Calogero model with the oscillatory potential.
Actually the relation between Calogero type models and FQHE was
known for a while  \cite{iso} however it was not exploited so far.

Therefore the natural question is if there exists a formulation
of FQHE in terms of the  twisted commutative gauge theory.
The natural candidate for the commutative gauge description of the
FQHE is commutative CS action considered in \cite{fr}.
In this
paper we will show how such commutative description can be formulated
providing the realization of the Morita equivalence in the context
of FQHE. This issue has 
been briefly discussed in \cite{barbon}. 
In what follows we will consider  models with a
finite number of electrons since precisely in this case Morita
duality can be formulated in a very explicit way. First, we consider
the FQHE on the disc and map it into the model on the cylinder
described by the trigonometric Calogero (Sutherland) model using the dualities
known in the context of the integrable many body systems \cite{dual}.
Then we use the YM formulation of the trigonometric Calogero model
\cite{gn1} to clarify the version of Morita equivalence in this case.

It is the degenerate case of a more general situation when the
FQHE is defined on the torus and therefore there is a finite number
of  electrons from the very beginning. The equivalent many-body
system is now the relativistic trigonometric Ruijsenaars model
which is known to allow a description as a commutative CS theory
on the dual torus \cite{gn2}. In this case  Morita duality is quite
transparent.

The NCCS description of FQHE yields the IIA brane realization
of the corresponding theory \cite{susskind2,bob} via the configuration
of D0, D2 and D6(D8) branes. In the context of D branes Morita duality
is nothing but T duality transformation. Therefore we shall
match the brane interpretation of FQHE with the T dual realizations
via the gauge theories. On the other hand we shall exploit the
brane realizations of the integrable systems known in the 
context of  SUSY gauge theories (see \cite{gm} for a review 
and references therein). 
In this paper we suggest new IIB type
description for the FQHE in terms of D1, D3, and 5 branes
based on the known realization
of the CS term in IIB type theory \cite{csbranes}. We develop
the IIB brane picture for  Jain and generalized 
hierarchies. 

Besides the T duality transformation mapping between 
two equivalent description of the given QH system 
there are additional duality which are consistent 
with RG flows between the members of the whole FQHE
family. These duality transformations constitute
the infinite discrete group $\Gamma_{U}(2)$ and
impose the restriction on the $\beta$ function in the
theory. Using the IIB picture we identify all
basic transforms from the duality group in the
brane terms. Moreover we conjecture the brane 
interpretation of the RG transitions between
the plateaux  as the motion of D3 branes 
in the five branes web. This conjecture appears
to be in the rough agreement with the existence 
of the critical point between two plateaux. 

Let us emphasize that we have branes of
different dimensions involved in our 
configuration. Therefore any physical 
phenomena concerning the Quantum Hall system
looks differently from the point of view 
of the gauge theories defined on the
worldvolumes of D1, D3 or (p,q) 5 branes. 
Namely, from D1's point of view one deals 
with the $\sigma$ model on the complicated
manifold determined by the brane configuration.
From the point of view of the theory on D3's 
we will consider the pure Chern-Simons theory
or Maxwell Lagrangian with CS term. Finally,
theory defined on the five branes provides
the d=6 or d=5 theory where some defects,
for instance monopole like configurations,
are represented by D1 and D3 branes.

The paper is organized as follows. In Section 2 we describe the
realization of the FQHE on the disc in terms of the
commutative YM on the cylinder with inserted 
Wilson line. In Section 3 we consider
the Morita dual description of FQHE on the torus 
using the formulation of the model with
the finite amount of electrons  in terms 
of the Ruijsenaars integrable many-body system. 
In Section 4 we consider duality transformations
relevant to the RG behavior of the FQHE systems 
and compare them with the dualities known in the
context of the Calogero-Ruijsenaars type models.
Section 5 is devoted to the
new IIB brane picture 
which is  applied for the Jain and generalized hierarchies.
In Section 6 we identify 
transformations from the duality group $\Gamma_{0}(2)$
of the RG flows as well as the particle-hole symmetry
in the brane terms. The picture of RG flows between
the two plateaux in the brane terms is conjectured. 
Section 7 contains some  conclusions and speculations.

\section{U(1) NCCS and YM on the cylinder}

\subsection{The Chern-Simons matrix model}

The starting point of our considerations is the realization of  U(1) NCCS
theory via  infinite matrixes $X_a$  \cite{susskind}
\beq
S=\int dt BTr \{ \epsilon_{ab}(\dot {X_a} +i[A_0,X_a])X_b +2\theta A_0 \}
\eeq
where the trace over the Hilbert space substitutes the integration
over the noncommutative plane in the matrix formulation. The averaged
density of electrons is
\beq
\rho=\frac{2\pi}{\theta}
\eeq
The coefficient in front of the CS action is related to the
filling fraction $\nu$ as follows
\beq
k=\frac{B\theta}{4\pi}=\frac{1}{4\nu}
\eeq
The action enjoys  gauge invariance and the Gauss law constraint
supports the noncommutativity of the coordinates
\beq
[X_1,X_2]=i\theta
\eeq

To consider the model with a finite number of electrons it was
suggested to use the finite gauged matrix model \cite{poly3}
\beq
S=\int dt BTr \{ \epsilon_{ab}(\dot {X_a} +i[A_0,X_a])X_b +2\theta A_0  -
B X_a^2 \} + \Psi^{+}(i\dot {\Psi} -A_0\Psi)
\eeq
where coordinates are represented now by $N \times N$ matrixes. The
additional fields $\Psi$ which are necessary
to be introduced take care of the boundary of the droplet and  transform
in the  fundamental of the gauge group U(N). The density of the
electrons in the model $\rho=\frac{N}{\pi R^2}$ coincides with
the infinite case since the radius of the droplet is
of order $\sqrt {N\theta}$.The excitations of the finite model perfectly
match the excitations known  in the FQHE \cite{Hellerman}.

It can be shown that after the resolution of the Gauss law
constraint the matrix model becomes equivalent to the rational
Calogero model 
\beq
H=\sum_{n=1}^N(1/2 p_n^2 +1/2B^2X_n^2) +
\sum _{n\neq m}\frac{\nu^{-2}}{(x_n -x_m)^2}
\eeq
The positions of the Calogero particles are the eigenvalues of $X_1$
while momenta can be identified with the diagonal elements of $X_2$.
Since the Hamiltonian of the Calogero model coincides with the
potential in the matrix model the states map between the two models.
In what follows we shall use the mapping to the many-body systems
to formulate the commutative gauge theory.

\subsection{Mapping to the trigonometric Calogero model}

To get the link with the gauge theory 
let us describe at the first step the map of the model above into the
Sutherland model with the Hamiltonian
representing a system of indistinguishable
particles on a circle $S^{1}_{R}$
of the radius $R$, interacting
with the pair-wise potential \cite{nikita}
\beq
U_{suth}(q) = {{{\xi}^{2}}\over{4R^{2} {\sin}^{2}
( {{q}\over{2R}} )}}
\eeq
This is a simple example of the dualities between the pairs
of the integrable many-body systems  \cite{dual} which can be
effectively described in terms of the Hamiltonian
reduction procedure.

From the Hamiltonian point of view the system has the phase
space:
\beq
M_{suth} = [ T^{*}\bigl( S^{1}_{R} \bigr)^{N} ] /{S}_{N}
\eeq
where $S_{N}$ is the $N$-th order symmetric group. The coordinates
in the phase space will be denoted as
$(p_{i}, q_{i})$ where $q_{i}$ is the
angular coordinate on the circle $S^{1}_{R}$
and $p_{i}$ is the corresponding momentum.
The Hamiltonian of the many-body system $H_{suth}$
has the natural form:
\beq
H_{suth} = \sum_{i=1}^{N}
{{p_{i}^{2}}\over{2}} +
\sum_{i \neq j} U_{suth}(q_{i} - q_{j})
\eeq
which is a well-known Sutherland model (see \cite{olper}
for the review on the Calogero-Sutherland models).

We shall present now the explicit map between
the Sutherland model and the Hamiltonian (6)
providing the description of the FQHE on the disc.
The phase space of the rational model is
\beq
M_{cal} =[ T^{*} {R}^{N} ] /{S}_{N}
\eeq
Hereafter we will extensively use the group like
description of the Calogero type models. The most effective
approach appears to be the Hamiltonian reduction
procedure.

Recall how the Hamiltonian reduction gives rise to the
systems discussed.
The main idea behind the construction is to realize the
(classical) motion due to the Hamiltonians
as a projection of the simple motion on a
somewhat larger phase space.
In the Sutherland case one starts with the symplectic
manifold $X_{suth} = T^{*}G \times {C}^{N}$ where $G = U(N)$ with the
canonical Liouville form
\beq
\Omega_{suth} = i Tr  \delta ( p \wedge \delta g g^{-1}  )
+{1\over{2i}} \delta v^{+} \wedge \delta v
\eeq
Here $p$ represents the cotangent vector to the group $G$.
We think of it as of the Hermitian matrix.
The manifold $X_{suth}$ is  acted on by $G$ by conjugation
on the $T^{*}G$ factor and in a standard way on $C^{N}$. The action
is Hamiltonian with the moment map:
\beq
\mu_{suth} = p - g^{-1}pg - v \otimes v^{+}
\eeq
One performs the reduction at the level of the moment map and takes
its quotient by $G$.
Explicitly, one solves the equation
\beq
p - g^{-1}pg - v \otimes v^{+} = -\xi 1
\eeq
up to the $G$-action. The way to do it
is to fix a gauge
$$
g = {\exp} ( {i\over{R}}  diag (q_{1}, \ldots, q_{N}) )
$$
and then solve for $p$ and $v$. One has:
$$
v_{i} = {\sqrt{\xi}}, \quad p_{ij} = R  p_{i} \delta_{ij} +
{\xi} {{1- {\delta}_{ij}}\over{e^{{i(q_{i}-q_{j})}\over{R}} -1}}
$$

As a result one gets the reduced phase space:
\beq
M_{suth} = [ T^{*} ( S^{1} )^{N} ] /S_{N}
\eeq
with the canonical symplectic structure
$$
\Omega_{suth}^{red} = \sum_{i} \delta p_{i} \wedge \delta q_{i}
$$
The complete set of functionally independent
integrals is given by:
\beq
H_{suth}^{(k)} = {1\over{R^{k+1}}}Tr p^{k+1}, \qquad k = 0, \dots, N-1
\eeq

The unreduced phase space of the rational Calogero system
has the form $X_{cal}=T^{*}g \times C^{N}$ where $g=Lie U(N)$.
The group acts on this manifold and upon the reduction on the level of the
moment map one obtains the reduced phase space mentioned above.The
set of independent Hamiltonians in the rational model amounts from
\beq
H^{k+1}=Tr(ZZ^{+})^{k+1}
\eeq
where $Z=P+iQ, (P,Q) \in Lie U(N)$.

The symplectic map between two models looks as follows. First,
elements from
$C^{N}$ get unchanged, coupling constants in two models
coincide while the map sends Z to the polar decomposition
\beq
Z=\sqrt {Bp} g
\eeq
where  p is supposed to be hermitian matrix with the non-negative
eigenvalues. The frequency of the oscillator in the rational model gets mapped
into the radius in the Sutherland model
\beq
B=R^{-1}
\eeq
and the quadratic Hamiltonian of the Calogero model is mapped to
the total momentum of the Sutherland model. Higher hamiltonians are mapped
as well, for instance quadratic Hamiltonian of the Sutherland model
is mapped in the quartic Hamiltonian in the Calogero system. This
correspondence survives at the quantum level and the spectrum of the
Hamiltonian of the Calogero model exactly coincides with the
spectrum of the total momentum in the Sutherland model.

\subsection{Sutherland system from 2D YM theory}

Let us show how the Lagrangian formulation of the Sutherland systems
amounts from the YM theory on the cylinder
\cite{gn1}.
Consider the central extension of the loop algebra
${\cal L} G$, where $G$ is a semisimple Lie algebra.
We take the cotangent bundle to the algebra as a bare phase space, namely
$(A,k;\phi,c)$, where $\phi$ - $G$
- valued scalar field , $c$ - the central element  ,
$A$ - the gauge field on the circle  and $k$ -
is dual to $c$.
A natural symplectic structure is defined
as follows
\beq
\Omega = \int Tr ( \delta \phi \wedge \delta A) + \delta c
\wedge \delta {\kappa}
\label{sympl}
\eeq
while the adjoint and coadjoint
action of the loop group ${\cal L}G$
on $\hat g$ and ${\hat g}^{*}$ reads
\beq
(\phi(\varphi) , c) \rightarrow  (g(\varphi) \phi(\varphi) g(\varphi)^{-1},
\int Tr (- \phi g^{-1}{\partial_{\varphi}}g) + c )
\label{adj}
\eeq

\beq
(A, {\kappa}) \rightarrow (g A g^{-1} + {\kappa} g {\partial_{\varphi}}
g^{-1},
{\kappa})
\label{coadj}
\eeq
This action clearly preserves the symplectic structure and thus defines a
moment map
$$
\mu : T^{*}{\hat g} \rightarrow {\hat g}^{*},
$$
which takes
$(\phi, c; A, {\kappa})$ to $( {\kappa} d \phi + [A, \phi] , 0)$.

Let us remind that the choice of the level of
the moment map has a simple physical meaning -
in the theory of the angular momentum just the sector of the
Hilbert space is fixed. In the gauge theory
one selects the gauge invariant states.
It is natural to consider an  element
$J$ from $g^{*}$,
which has the maximal stabilizer different from the whole
$G$. It is easy to show that the representative of the
coadjoint orbit of this element has the following form

\beq
J_{\nu} = {\nu} \sum_{\alpha \in \Delta_{+}}(
e_{\alpha} + e_{-\alpha}),
\label{mom}
\eeq
where  $\nu$ is some real number,
$e_{\pm \alpha}$  are the elements of nilpotent subalgebras $n_{\pm} \subset
g$, corresponding to the roots $\alpha$, and ${\Delta}_{+}$ is
the set of positive roots.
Let us denote the orbit of $J_{\nu}$ as ${\cal O}_{\nu}$.  When
quantizing ${\cal O}_{\nu}$ we get a
$R_{\nu}$ representation of $G$.
For generic $J \in g^{*}$ denote as $G_{J}$ the stabilizer $J$ of
the orbit ${\cal O}_{J}$, that is ${\cal O}_{J} = G / G_{J}$.  Therefore the
proper orbit in the affine case ${\cal O}_{\nu}$

\beq
\mu = ({\cal J}[{\mu}],0) : {\cal
J}[{\mu}]({\varphi}) = \delta({\varphi}) J_{\nu} \label{affmom}
\eeq
is
finite dimensional.

To perform a reduction one has to resolve the moment map
which has the meaning of a Gauss law in the affine case
\beq
{\kappa} \partial_{\varphi} \phi + [A, \phi] =  J_{\mu}({\varphi}) =
\delta({\varphi}) J_{\nu}.
\label{redeq}
\eeq

It is useful to resolve the constraint in the following manner.
First, we use the generic gauge transformation $\tilde g({\varphi})$ to
make
$A$ to be a Cartan subalgebra valued one form $D$.
We are left with the freedom to use the constant gauge transformations
which do not affect
$D$.  Actually the choice of $D$ is not unique and is parameterized by the
conjugacy classes of monodromy
$\exp (
\frac{2\pi}{\kappa} D) \in {\bf T} \subset G$.
Let us fix the conjugacy class
denoting as ${\sf i} x_{i}$ the elements of the matrix $D =
iX$.  Let us decompose the $g$ - valued function $\phi$ on the $S^{1}$ on
Cartan $P(\varphi) \in t$ and nilpotent ${\phi}_{\pm}({\varphi}) \in
n_{\pm}$ components.  Let ${\phi}_{\alpha} = <{\phi}, e_{\alpha}>$, then
(\ref{redeq}) reads:
\beq
{\kappa}\partial_{\varphi} P = \delta({\varphi}) [ J_{\nu}^{g}
]_{\gamma}
\label{careq}
\eeq

\beq
{\kappa}\partial_{\varphi}{\phi}_{\alpha} + <D, {\phi}_{\alpha}> =
\delta({\varphi}) {[ J_{\nu}^{g} ]}_{\alpha},\\
\label{rooteq}
\eeq
where $J_{\nu}^{\tilde g}$ - $Ad_{{\tilde
g}(0)}^{*}(J_{\nu})$,
$[J]_{\gamma}$ denotes the Cartan part of $J$ and
$[J]_{\alpha} = <J,e_{\alpha}>$.

From (\ref{careq}) one gets $D = constant$ and
$[J_{\nu}^{g}]_{\gamma} = 0$.
Therefore we  can twist back
$J^{g}_{\nu}$ to $J_{\nu}$ .
Moreover, (\ref{rooteq}) implies, that (if $\varphi \neq 0$ )
can be presented ${\phi}_{\alpha}({\varphi})$ in the following form:

\beq
{\phi}_{\alpha}({\varphi}) =
\exp ( - \frac{\varphi}{\kappa} <D, {\alpha}> ) \times M_{\alpha},
\label{rootres}
\eeq
where $M_{\alpha}$ is locally constant element in
$g$.  It is evident that $M_{\alpha}$ jumps, when ${\varphi}$ goes through
$0$. The jump is equal to
\beq
[\exp ( - \frac{2\pi}{\kappa} <D, {\alpha}> )
- 1] \times M_{\alpha} = [J^{g}_{\nu}]_{\alpha}.
\label{jump}
\eeq
Finally  the physical degrees of freedom
are
$\exp(-\frac{2\pi i}{\kappa}X)$ and $P$ with the symplectic structure
\beq
\omega = \frac{1}{2 \pi i} Tr ( \delta P \wedge \delta X )
\label{redsym}
\eeq
and
$$
{\phi}_{\alpha}({\varphi}) = {\xi} \frac{\exp ( -
\frac{i\varphi}{\kappa} <X, {\alpha}> )} {\exp ( - \frac{2\pi i}{\kappa} <X,
{\alpha}> ) - 1}.
$$

Collecting together the term yielding the Poisson structure
moment map with the lagrangian multiplier $A_{0}$
and the second Casimir
\beq
{\cal H}_{2} = \frac{1}{4\pi} \int_{S^{1}} d{\varphi} <{\phi},
{\phi}> ,
\label{2cas}
\eeq
we immediately recognize the 2d YM theory in the
hamiltonian formulation with the additional Wilson line included.
On the reduced manifold we obtain the Sutherland
system
with the hamiltonian
\beq
H_{2} = -\frac{1}{2} Tr P^{2} +
\sum_{\alpha \in \Delta_{+}} \frac{{\xi}^{2}}
{sin^{2}<X,{\alpha}>} ,
\label{redham}
\eeq
where $\Delta_{+}$
denotes the positive roots of the group
$g$. For instance for  $G = SU(N)$   we get the hamiltonian with the pairwise
interaction

$$
V_{ij}^{A} = \frac{\xi^{2}}{sin^{2}(x_{i}-x_{j})}
$$

Summarizing:  in this Section we presented the two step map
of U(1) NCCS theory on the disc into the commutative
SU(N) YM theory on the cylinder 
with inserted Wilson line 
yielding the degenerate version of the Morita
duality. One should not be confused that 
$D=3$ CS theory  is related to  $D=2$ YM
theory on the cylinder. The point is that 
YM theory on the cylinder can be derived 
from the G/G gauge $\sigma$ model or equivalently
CS theory in the limit of  large level k. On the 
other hand since one of the radii of the torus where CS
is defined on is $R\propto k^{-1}$ the limit of large k
corresponds to the theory in one dimension less.
This  will be more clear in the next section
where the relation to the Ruijsenaars model
will be discussed.  

\section{Ruijsenaars model and NCCS theory on the torus}

In this section we will generalize the Sutherland model to the so called
trigonometric Ruijsenaars model and will argue that this model provides
the
finite dimensional counterpart for the FQHE on the torus. Naively
on the torus both momenta and coordinates have to be periodic and
precisely Ruijsenaars model yields the appropriate phase space. It will
be shown that this dynamical system has the moduli space of the flat
connections on the torus with one marked point as  phase space
and the relation coming from the fundamental group gets mapped
into the noncommutativity relation between coordinates on the dual torus.
The rank of the gauge group G will coincide with the number of electrons 
in the FQHE.

In the Hamiltonian reduction approach we start with the cotangent
bundle to the loop group \cite{gn2}
$\hat G$.
$(g:S^{1} \rightarrow G, c \in U(1) ; A \in \Omega^{1}(S^{1}) \otimes g^{*},
{\kappa}
\in {\bf R})$.
The group acts as
$$
g \rightarrow h g h^{-1}, \; A \rightarrow h A h^{-1} + {\kappa} h {\partial}
h^{-1} 
$$
The generalization of the Gauss law is
\beq
\mu (g,c; A,\kappa) = (g A g^{-1} + {\kappa} g {\partial} g^{-1} - A,
0).
\label{momgru}
\eeq

The level of the moment map looks as follows
\beq
\mu (g,c; A,{\kappa}) = {\sf i} \nu (\frac{1}{N}
Id - e \otimes e^{+}) \delta (\varphi).
\label{momlev}
\eeq
As before, a general gauge transformation can be used to
reduce
$A$ to the diagonal form $D$
modulo the affine Weyl group action.

The moment map now becomes
\beq
g D g^{-1} + {\kappa} g d g^{-1} - D =
{\sf i} \nu (\frac{1}{N} Id - f \otimes f^{+})\delta(\varphi)
\label{mastereq1}
\eeq
where $f $ - vector  with the unit norm $<f,f>
= 1$.
$f \in {\bf R}^{N}$,

$$
g = \exp ( \frac{\varphi}{\kappa} D) G(\varphi)
\exp( -\frac{\varphi}{\kappa} D) ;
\partial_{\varphi} G = - \frac{J}{\kappa} G \delta(\varphi),
$$
where $J = {\sf i} \xi ( \frac{1}{N} Id - f \otimes f^{+} )$.
It is useful to introduce the monodromy of the connection
$D$: $Z = \exp ( - \frac{2\pi}{\kappa} D) =
diag(z_{1}, \dots, z_{N})$,  $\prod_{i} z_{i} = 1$, $z_{i} = \exp
(\frac{2{\pi}i q_{i}}{\kappa})$ with the boundary condition:
\beq
{\tilde G}^{-1} Z {\tilde G} = \exp(\frac{2\pi J}{\kappa}) Z
\label{commutant}
\eeq
$ {\tilde G} = G(+0)$.

The solution to the equation can be written in terms of the
characteristic polynomial
$P(z)$ of the matrix  $Z$, ;
$$
P(z)=\prod_{i} (z - z_{i}).
$$
Let
$$
Q^{\pm}(z) =
\frac{P({\lll}^{\pm 1}z) - P(z) }{({\lll}^{\pm N}-1)z P^{\prime}(z)},
$$
where $\lll = e^{\frac{2 \pi i\nu}{N \kappa}}$ .Then if $\lll \to 1$,
rational functions $Q^{\pm}(z)$ tend to $\frac{1}{N}$.
In this notations  $\tilde G$ looks as:

\beq
{\tilde G}_{ij} = - \lll^{-\frac{N-1}{2}}
\frac{{\lll}^{-N}-1}{{\lll}^{-1}z_{i}-z_{j}}
e^{i\theta_{i}}(Q^{+}(z_{i})Q^{-}(z_{j}))^{1/2}
\label{Lax}
\eeq

$$
=e^{i\theta_{i}-\frac{\pi i}{\kappa} (q_{i}+q_{j}) }
\frac{sin(\frac{\pi\nu}{\kappa})}{sin(\frac{\pi(q_{ij}-
\frac{\nu}{N}}{\kappa})}
\prod_{k \neq i, l \neq j}
\frac{sin(\frac{\pi{q_{ik} +
\frac{\nu}{N}}}{\kappa})}{sin(\frac{{\pi}q_{ik}}{\kappa})}
\frac{sin(\frac{{\pi}{q_{il} -\frac{\nu}{N}}}{\kappa})}{sin(
\frac{{\pi}q_{il}}{\kappa})}
$$
where
$\theta_{i}$
can be identified with the momenta of particles with coordinates
$q_{i}$.
the reduced symplectic structure is
$\sim \sum_{i} d\theta_{i} \wedge dq_{i}$.

The natural gauge invariant Hamiltonian is
$$
H_{\chi} = \int d{\varphi} {\chi}(g).
$$\beq
H_{\pm} = \sum_{i} (e^{i \theta_{i}} {\pm} e^{- i\theta_{i}})
\prod_{j \neq i} f(q_{ij})
\label{ruuham}
\eeq
with the function $f(q)$:
$$
f^{2}(q) = [ 1 -
\frac{sin^{2}
({\pi}{\nu}/{\kappa} N)}
{sin^{2}({\pi}q/{\kappa})}],
$$
which yields the Ruijsenaars model , while in the limit
$\kappa \to \infty$ we return back to the Sutherland one.

To fix the corresponding gauge theory we take  $\int pdq$  and add
the moment map equation as a constraint.
The resulting action  is nothing but the action of the   $G/G$  sigma model
with the Wilson line included \cite{gn2}.
The equivalent representation involves Chern-Simons theory
on 
$X = I \times T^{2}$ with the action:
\beq
S_{CS} = \frac{i {\kappa}}{ 4\pi}\int_{X} Tr (A \wedge dA +
\frac{2}{3} A \wedge A \wedge A).
\label{CS}
\eeq
The phase space now is the moduli space of the flat connections
on the torus with the marked point where
the peculiar coadjoint orbit 
is inserted. The latter modifies the path integral as
$$
D A <v_{1}| T_{R}(P \exp \int A ) |v_{2}> \exp( - S_{CS}(A) ).
$$

The monodromy around the marked point
$U$  is specified by the highest weight $\hat h$ of the representation
$R_{\nu}$  as
$ U = \exp ( \frac{2\pi {\sf i}}{{\kappa} + N} \mbox{diag} ( {\hat
h}_{i}))$.

Let 
$g_{A}, g_{B}$ be the  monodromies around the cycles on $T^{2}$,
and  $g_{C}$  the monodromy around the marked point.Then the monodromy condition on
$T^{2}$ is:
$$
g_{A} g_{B} g_{A}^{-1} g_{B}^{-1} = g_{C}
$$
This relation becomes the condition of the noncommutativity of
the coordinates on the dual torus. It can be considered 
as generalization of the noncommutativity relation previously discussed 
in the FQHE context on the disc and on the cylinder.

The natural gauge invariant Hamiltonian
$cos( \frac{2\pi{\sf i}}{{\kappa} + N} {\partial}_{q})$
amounts to the difference operator with the spectrum
$e^{2 {\pi}{\sf i}n q}$

$$
E_{n} = cos (\frac{2\pi n}{{\kappa} +N}).
$$
The spectrum reads
$$
\sum_{i} E_{n_{i}} ,
$$
where we use the invariance
$E_{n} =
E_{n +{\kappa} + N}$ and symmetry:
\beq
({\kappa} + N )>
n_{N}> \dots > n_{i} > \dots > n_{1} \geq 0 \label{restr1}
\eeq
\beq
\sum_{i} n_{i} < ({\kappa} + N)
\label{restr2}
\eeq
It would be important to compare the excitation
spectra in the Ruijsenaars model and the
FQHE on the torus.

Now we are ready to formulate the Morita duality in the
context of FQHE. Morita duality acts on the
gauge bundle on the torus as SO(2,2,Z)
transformation \cite{morita}. Generically it maps
the nonabelian gauge field on the
noncommutative torus $T^2_{\theta}$ to the
abelian or nonabelian field on the dual
torus  $T^2_{\tilde{\theta}}$. The
parameters of noncommutativity are related
by the SO(2,2,Z) transformation
\beq
\tilde{\theta}=\frac{a+b \theta}{c-d \theta}
\eeq
The rank of the gauge group as well as the
flux $\int F=m$ get transformed under the duality
and we shall be interested in the transformation
which brings the  abelian noncommutative
bundle to the twisted nonabelian one.
Note that the Morita duality is nothing but
the T duality transform in the stringy context.

Under the duality transformation the rank and the
flux get interchanged therefore starting with U(1)
noncommutative theory we end  with U(N) twisted
abelian bundle on the
dual torus where N is the number of electrons.
It is necessary to check what happens with the
level k in the CS action. It was shown in \cite{barbon}
that the effective level 
in CS action doesn't change under the Morita 
transformation. Using this fact and the consideration
above we conclude that two T dual descriptions 
of FQHE fit perfectly.

\section{Duality in FQHE and in  Calogero type  systems}
\subsection{Dualities in FQHE}
In this section we discuss the
dualities known in the context of the Quantum Hall
Effect and compare them 
with the dualities in the 
Calogero-Ruijsenaars 
type model as well with 
dualities in the context of 
supersymmetric gauge theories. 
In the previous section we considered the Morita 
duality which is the T duality transformation. 
It can be used in a given  QH system and corresponds
to two equivalent descriptions of this particular
system.

The key difference with the T duality
transformation which can be applied to a given system
is that other dualities can be naturally 
applied to the whole bundle of the QHE systems
over the parameter space. More exactly 
these symmetries can be considered as the
symmetries of the RG flows between the
FQHE platoeaux. To formulate these symmetries
qualitatively it was suggested that the  
discrete group  acts on the complexified 
conductivity  \cite{kivelson}
\beq 
\sigma=\sigma_{xy} +i\sigma_{xx}
\eeq

The basic transformations which yield
the duality group are the Landau level 
addition (L);
\beq
\sigma_{xy}(\nu +1) \rightarrow \sigma_{xy}(\nu) +1 
\qquad \sigma_{xx}(\nu +1) \rightarrow \sigma_{xx}(\nu)
\eeq
the flux attachment transformation(F);
\beq
\sigma^{-1}_{xy}(\frac{\nu}{2\nu+1}) \rightarrow \sigma^{-1}_{xy}(\nu) +2 
\qquad \sigma_{xx}(\frac{\nu}{2\nu +1}) \rightarrow \sigma_{xx}(\nu)
\eeq
and the particle-hole transformation
\beq
\sigma_{xy}(1-\nu ) \rightarrow 1- \sigma_{xy}(\nu)  
\qquad \sigma_{xx}(1- \nu ) \rightarrow \sigma_{xx}(\nu).
\eeq
On the plateaux these transformations can be expressed 
purely in terms of the filling factors since 
there $\sigma_{xx}=0$. The powers of the transformations
L and F amount to the discrete group of the
infinite order  $\Gamma_{U}(2)$. It also appeared
that the particle-hole symmetry which is the
outer automorphism of the  $\Gamma_{U}(2)$ group
amounts to the additional constraints on the 
structure of the RG flows \cite{lr}.   
The
typical parameter of the RG flow could be the external
magnetic field.

Using the different elements from the 
duality group one could generate arbitrary
QH state from some fixed one. For instance,
starting from $\sigma=1$ one could obtain 
the integer QH states with $\sigma=m$ by the
transformation $L^{m-1}$, the Laughlin states
$\sigma=\frac{1}{2m+1}$ by the action of $F^{m}$
and the Jain states $\sigma=\frac{p}{2mp+1}$ 
by the action $F^{m}L^{p-1}$.

Many properties of the QHE system 
just follows from the consistency of the RG flows 
with the  duality group. For instance 
the semi-circle law \cite{bdd}, universality of the transition 
conductivities as well as selection rules 
for the transitions between plateaux 
\cite{dolan} follow
from the duality property. The universal
critical points which map to themselves 
under the duality group are predicted. These
points are located at $2\sigma_{crit}=1+i$
and their images under the duality group. The
duality group, for instance, predicts that the transition
between Hall plateaux with the fractions $p_1/q_1$
and $p_2/q_2$ is possible only if 
$$
|p_1q_2 - p_2q_1|=1.
$$  

The discrete group provides some prediction
concerning the complex $\beta$ function 
of the theory. It can be proven \cite{bl}
that the $\beta$ function 
of the theory introduced in \cite{scaling}
should obey the
following equation
\beq
\beta(\gamma(\sigma),\gamma (\bar{\sigma}))=
\frac{\beta(\sigma, \bar{\sigma})}{(c\sigma +d)^2}
\eeq
where
\beq
\gamma(\sigma)= \frac{a\sigma +b}{c\sigma +d}
\eeq
with even c and $ad - bc=1$.
It is also known  that in the weak
coupling regime the $\beta$ function behaves as
\beq
\frac{d\sigma}{d logL}= -\frac{i}{2\pi Im\sigma} +
higher \ loops + instanton  \ contribution
\eeq

Let us discuss now the analogies with the duality
group known in the context 
of SUSY YM theories. First,  in our
opinion the analogy with N=2 SUSY theories 
sometimes mentioned in the literature is incorrect.
The point is that the N=2 SUSY YM, contrary 
to FQHE, enjoys the huge moduli
space therefore we could connect the nontrivial RG
flows along the moduli space with the duality
group in a given theory. The coordinate on the moduli space
provides some RG scale and one could investigate 
the dependence on this coordinate. 

The proper 
SUSY system to compare with is the N=1 theory
which has no moduli space unless the fundamental
matter is introduced. This theory similar to the 
QHE case has all loop contributions to the
$\beta$ function nevertheless at least perturbatively
it can be calculated exactly \cite{nsvz}.
As for the duality group the relevant 
duality is the one of the RG flows
between the theories with the different $N_c,N_f$.
The famous Seiberg duality \cite{seiberg} 
between the $SU(N_c)$ 
theory with $N_f$ flavors and $SU(N_f - N_c)$
with $N_f$ flavors and additional superpotential is 
the most elaborated example. The mirror symmetry
is its d=3 counterpart. 
We will return to this analogy when the IIB brane 
picture for FQHE will be considered.

\subsection{On dualities in Calogero type systems}
Since the  Calogero system describes the
FQHE for a finite number of electrons 
it is reasonable to look for the  map 
between the
duality known in Calogero context 
to the one in FQHE. The first symmetry
to be discussed is the mapping of the 
Calogero coupling constant $\nu$ to
$\nu^{-1}$ \cite{calogero}. Since 
the coupling constant $\nu$ in the noncommutative
U(N) CS formulation is $N/k$ where N is the rank of the
gauge group and k is the level, the duality is
the version of rank-level duality. It can be
also treated as a version of 3d mirror transformation
\cite{kapustin}. 

To discuss the duality property let us  
proceed as follows. 
Let us represent the trigonometric Calogero Hamiltonian in terms
of the creation-annihilation operators
\begin{equation}
H= \sum_{n}((1-\nu)na_{n}^{+}a_n +\nu Na_{n}^{+}a_n) +
\sum_{mn}(\nu a_{m}^{+}a_{n}^{+}a_{n+m} +a_{m+n}^{+}a_{m}^{+}a_{n})
\end{equation}
To get the dual picture introduce new variables $\nu b_{n}^{+}=-a_{n}^{+}$
and $b_n= -\nu a_n$. The Hamiltonian in terms of the new variables 
is identical to the previous one  upon $\nu$ gets substituted by $\nu^{-1}$
and N by -$\nu N$. The duality actually maps quasiparticles to
quasiholes in the spectrum of excitations. The mapping of the coupling
constant reflects the fact that the k quasiparticles have the opposite
flux to the k$\nu$ quasiholes.

One more duality has a rather simple 
manifestation. At the quantum level 
the Calogero Hamiltonian depends 
only on the product g(1-g) where 
the Calogero coupling was identified
with the inversed filling factor. Therefore
the evident duality $g \rightarrow 1-g$
corresponds the transformation 
$\nu^{-1} \rightarrow 1 - \nu^{-1}$.

It is known that the Calogero-Toda  
type systems  also are  relevant for the
exact Seiberg-Witten solution to N=2 theory 
(see \cite{gm} and the references therein).
Namely the periodical Toda lattice 
governs the nonperturbative dynamics in the pure
N=2 SUSY gauge theory while 
the elliptic Calogero system provides
the exact solution to the N=2 theory
with the massive adjoint hypermultiplet. 
However as we have mentioned above the
analogy with N=1 theories is more relevant.
The relation of Calogero system to N=1 theory
looks as follows. Now only the static 
configurations of the integrable many-body systems
are relevant and they are mapped to the vacuum states
of  N=1 theories. One more remark is in order here.
The point is that the coupling constant g  of Calogero
system (actually its Toda limit) can be identified 
with $\Lambda_{QCD}$ scale. From this point of view
the duality $ g \rightarrow  g^{-1}$  
is mapped into the relation 
$\Lambda_{QCD} \rightarrow \Lambda^{-1}_{QCD}$. 
The situation strongly resembles
the relation between the nonperturbative scales in
the Seiberg dual electric-magnetic pair
$$
\Lambda_{el}\Lambda_{magn}=const.
$$

The similarities between the RG behavior
of the N=1 SUSY theories and FQHE  
and the role playing the Calogero type
systems for both of them could provide the
additional insights on the problem.
Although the  theories are essentially
different they could  manifest  similar
universality classes with respect to the RG flows.
This relation also 
implies the existence of the   many-body system 
counterpart of
the S-duality known in SYM context.
This problem was questioned in \cite{dual} 
where it was shown that S duality acts
on the spectral curve of the integrable 
system as the modular transformation. 
Therefore from this point of view it is
necessary to recognize the spectral curve
of the corresponding integrable system
in the FQHE. However since we have 
discussed only trigonometric models 
the whole machinery of S duality can not
be used here. Nevertheless in what follows
we have found some interpretation of S-duality
transformations in the IIB brane picture 
and the brane interpretation of the RG will
be conjectured.

\section{Brane picture}
\subsection{IIA picture}
Let us turn to the brane description of FQHE.
Below we briefly review the IIA picture 
for the Quantum Hall system considered in
\cite{susskind2,bob,saidi}.

IIA picture for FQHE  according to \cite{susskind2,bob}
looks as follows. 
In the picture of Susskind et.al. one considers the D2 brane 
in the background of k D6 branes to get the filling
factor $\nu = k^{-1}$. The WZ terms in the D2 brane 
amount to the 
CS action at the level k   on D2 worldvolume. 
Therefore the natural gauge theory on the D2 
worldvolume is the Chern-Simons-Maxwell one.
Since in d=3 the YM coupling constant is
dimensionful it can be removed by the appropriate
scaling limit.
The noncommutativity in the remaining CS action
follows from the D0 branes. Electrons are represented 
by the ends of the strings connecting D6 branes and the D2 brane.

A more elaborate picture was obtained in \cite{bob} 
where the system of D2 and D8 branes is considered
in the massive IIA theory in the constant B field background.
The CS term is induced on D2 branes due to the 
D8 branes. The point is that in the massive IIA theory
in the NS B field background the RR rank-two 
field is generated as well
and plays the role of the magnetic field. 
In the B field the nontrivial density of D6 branes
on k D8 branes $ \rho_6 \propto kB$ is induced as well as 
density of D0 $\rho_0 \propto NB$ on N D2 branes. The filling
fraction in this case is $\nu = Nk^{-1}$.
Now electrons 
are represented by D0 branes while the ends of the strings
connecting D2 and D6 branes play the role of quasiparticles.
It is known that in the massive IIA theory a fundamental
string ending on any D brane carries 1/k units of 
the D0 charge which is consistent with QHE picture.

Let us explain how the Morita or T duality looks 
in IIA brane picture.
Following \cite{bob} we   consider one D2 brane
wrapped around the torus, N D0 branes localized on the torus
and representing the electrons and D8 branes generating
the Chern-Simons term $ k_{nc}$ on D2 worldvolume. According to 
the general
rules the noncommutativity can be read off from the brane picture

\beq
\frac{\theta}{Vol}=\frac{N_{D2}}{N_{D0}}
\eeq
and fits perfectly with identification as the
inverse density of electrons. Since in this description we consider
the gauge theory on the single D2 worldvolume
the theory is abelian.

After the T duality transformation along the torus
D2 and D0 get interchanged and
we obtain the nonabelian U(N) theory on the dual
torus with the unit D0 charge. Now
dimensionless $\theta$ is integer
and has to be treated as the twist.
Finally the T duality transformation applied
to D8 branes amounts to  D6 branes 
and the corresponding strings connecting D6 and
D2 branes induce  $k_{c}$ on  worldvolume of N D2 branes.


There was an attempt to
recognize the Jain and generalized hierarchies in the IIA picture
\cite{saidi}. 
In the Jain hierarchy the 
effective pinning of flux to the electrons 
corresponds to the bound state of the D0 branes 
on D2 brane
with 2m fundamental strings. Such composites
renormalize the magnetic field and the resulting
picture for the composite objects corresponds 
to the integer QHE. The second integer p 
corresponds to the number of D2 branes since 
it provides the rank of the group. More
general hierarchies involve more general 
configuration of D6 branes.

\subsection{IIB picture}

Let us suggest a new IIB interpretation 
of brane configuration for FQHE. The key point 
we are going to exploit is the IIB interpretation
of CS term \cite{csbranes}. To this aim consider 
the system of N D3  
branes with worldvolume coordinates $(0126)$
stretched between NS5 brane with coordinates $ (012345)$
and a (p,q) 5 brane with the same worldvolume 
coordinates. Note that both types  of 5 branes can be 
derived from the M5 brane in M theory assuming
the torus topology for two coordinates, say $x_2,x_{10}$.
In the IIB case NS5 branes amount from M5 brane 
with $(012345)$ worldvolume while (p,q) branes
amount from the M5 brane wrapping $x_2$ coordinate
q times and $x_{10}$ coordinate p times. It is also
known that D3 branes come from the M2 branes 
in M theory picture. We shall assume that 
D3 branes are stretched between 5 branes separated
by a distance $L_6$ in $x_6$ coordinate. The YM
coupling constant in the gauge theory on D3
branes is defined as 
\beq
\frac{1}{g_{YM}^2}=\frac{L_6}{g_s}
\eeq
where $g_s$ is the string coupling constant.

Recall now the generation of CS term in
such brane configuration \cite{csbranes}
The effective action in the gauge theory 
involves the interaction term
\beq
\int a(x) F \wedge F
\eeq
for the 
axion field $a(x)$. The arguments 
concerning the point-like instantons
imply that the different 
values of the axion field on
the 5 branes  yield the Chern-Simons 
term in the gauge theory 
on D3 branes at the level k=p/q. 

Another possible explanation
of the appearance of CS term comes from
the consideration of the boundary 
conditions for D3 brane 
$$
\partial_{\mu}A_6-\partial_{6}A_{\mu}=0 \qquad x_6=0 
$$
$$
\partial_{\mu}A_6-\partial_{6}A_{\mu} +
\epsilon_{\mu \nu \lambda}\frac{g_{4}^2 p}{4\pi q} 
\partial_{\nu}A_{\lambda} =0 \qquad x_6=L
$$
These boundary conditions arise when one takes
the variation of the action. If we write down
the d=4 YM action in three-dimensional terms
and take the variation over $A_{\mu}$ the
arising surface terms 
\beq
\int d^3 x \delta A_{\mu}( \partial_6 A_{\mu}- 
\partial_{\mu}A_6)
\eeq
have to be compensated. This aim can be achieved
by adding the CS term whose variation exactly
compensate the boundary term coming from the
YM action. 
In what follows we shall assume that the  standard YM term 
is small due to the closeness of
five branes.

Let us emphasize that in the Quantum Hall system 
there are no natural moduli spaces therefore we 
should forbid them by the brane configuration.
The induced CS term decreases the amount of SUSY
therefore D3 branes or M2 branes in M theory
picture can take only finite number of positions
corresponding to the finite number of the ground
states. Therefore the dangerous Coulomb branch 
is absent.

To make the
CS theory noncommutative let us assume that there are D1 
branes melting along D3 brane and hence  producing the
space noncommutativity. The noncommutative 
parameter for N=1 is the inverse density $\rho$ of D1 
branes $\theta = \rho^{-1}$. 
Note that we assume that D1,s are extended in 
$x_6$ direction therefore they represent
particles in 2+1 theory whose masses are
proportional to the distance between 5 branes.
Since  
noncommutativity can be identified 
with the inverse electron density $\theta = \rho_{e}^{-1}$
we have to identify electrons with D1 branes.
The quasiparticles 
and quasiholes correspond to the fundamental strings between 
D3 brane and (p,q) brane with different orientations. 
Since the fundamental string can not
end on NS5 there are no such additional states.

Let us now discuss the choice of 5 branes which
provide the different filling factors in QH system. 
The general formulae for the CS level for a 
pair of (p,q) 5 branes \cite{csbranes}
looks as follows
\beq
k=\frac{p_1q_2 - p_2q_1}{q_1q_2}
\eeq
It is clear that the simplest 
nontrivial choice involves
the NS5 and (2m+1,1) 5 brane and amounts to the 
level $k=2m+1$  corresponding to  Laughlin state with
 filling fraction  $\nu = 1/(2m +1)$ ( see Fig.1 b)

However this realization of the Laughlin state
is not unique. It can be derived 
if the additional ``flavors'' are added
to the configuration discussed above. 
To this aim we can add m additional semiinfinite D3 branes 
extended along $x_6$ coordinate or m 
additional D5 branes with worldvolumes $(012789)$.
Such additional branes usually represent the
matter in the fundamental in the gauge 
theories on the D brane worldvolumes. 
Consider now NS5 brane, (1,1) 5 brane 
and m additional massive flavors (see Fig.1 a). 
Then
assuming that the additional matter in fundamental is heavy
the desired renormalized CS level 
corresponding to the Laughlin state could be derived
after the matter is integrated out. It is 
this picture that will be generalized to the Jain hierarchy
case.  
 
\begin{figure}
\begin{center}
\epsfxsize=5in
\epsfbox{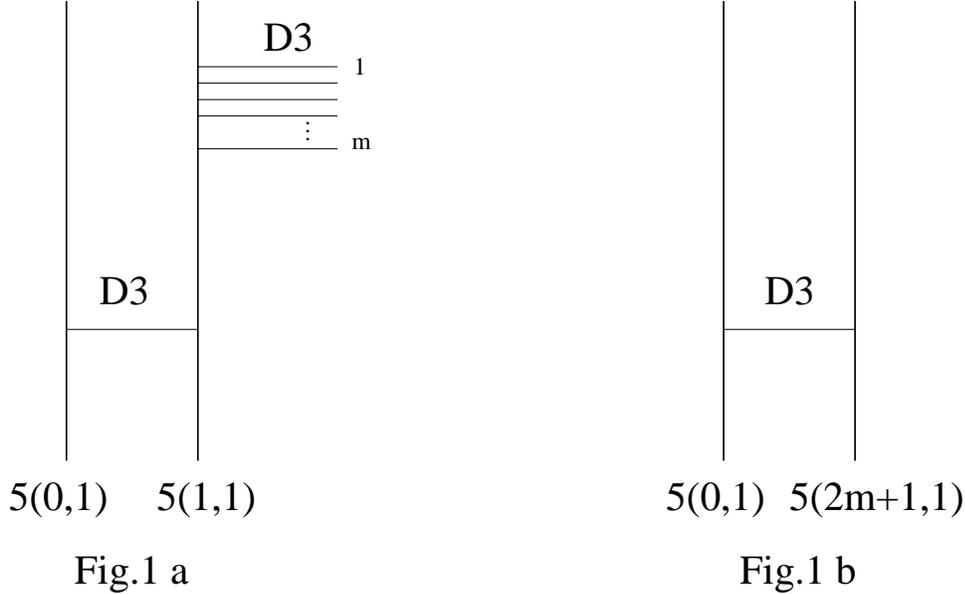}
\caption{Two IIB brane realizations of Laughlin states with $\nu =
\frac{1}{2m +1}$ with (Fig.1 a) and without (Fig.1b) auxiliary fermions}
\end{center}
\end{figure}

Let us briefly comment on the T duality transform which 
maps IIA into IIB picture. Assume that 
we perform T duality transform along  the 
coordinate which is effectively compact. Remind that
IIA picture involves D0,D2 and D6 branes. After 
T duality transform our IIB picture 
amounts just to  this brane content and D1 branes
in D3 branes get transformed into D0 branes  in D2 branes.
NS5 branes get unchanged under T duality 
while (p,q)5 branes amount into D6 branes.

One more remark is in order.
The IIB picture  is adequate to reproduce the phase transition
in level k recently found  in \cite{witten}. Namely
it was shown that  SUSY 
in the theory with the mass gap is spontaneously broken
at $k<N/2$ in N=1 SUSY case and $k<N$ in $N=2$ SUSY
case. Since our brane configuration 
presumably respects  N=2 SUSY 
we could expect  
that something happens with the stability 
of the brane configuration and therefore 
with the ground state of QH system at
the corresponding filling factors.
On the other hand we know from \cite{susskind}
that the filling fraction is just the ratio of
the rank to the level so the phase transition
in CS theory happens when we go 
through the $\nu=1$ state.

\subsection{Jain hierarchy and IIB brane picture}
Let us turn now to the IIB brane description
of the Jain hierarchy.
We will try to recognize the ``response model'' from 
\cite{kogan} in the brane terms. The Jain hierarchy
is fixed by two integers m and p which 
yield the following filling factor
\begin{equation} 
\nu = \frac{p}{2mp+1}
\end{equation}

The standard interpretation claims that
2m units of the magnetic flux are pined to
each electron. The resulting composite 
fermions feel the effective magnetic
field $B_{eff}=b - 4\pi m \rho_{e}$.
Therefore the effective filling fraction 
for the composite fermions reads as 
$\nu_{eff}=\frac{2\pi \rho_{e}}{B_{eff}}=p$
and hence the fractional quantum Hall states 
for the Jain filling factor can be related
with the integer quantum Hall states 
of the composite fermions.

In \cite{kogan} it was suggested to interpret
the Jain hierarchy as the result of the
perturbative renormalization of the integer
QHE by the auxiliary heavy fermions. It goes
as follows. First one considers the
$U(1)^p$ effective field theory description
with Chern-Simons action
\beq
S_{cs}(K)= K_{ij}\int d^3 x A^{i}dA^{j}
\eeq
where $i,j=1,....,p$. The filling factor is defined
by the inversed matrix K
\beq
\nu = Q^i K^{-1}_{ij}Q^j
\eeq
where Q is the charge vector.
To get the integer QHE
the matrix of CS terms in the $U(1)^p$ theory 
is of the following form
\beq
K^0_{ij}=\delta_{ij}
\eeq
Let us assume that there are auxiliary 
heavy  fermions interacting
with the external magnetic and CS fields  
and filling m first Landau levels 
with respect to the external magnetic field. If we 
integrate out the heavy fermions the CS matrix
gets renormalized from the loop correction
and acquires the following form
\beq
K_{ij}= K^0_{ij} + 2m (1)_{ij}
\eeq
where $(1)_{ij}$ is the matrix with all unit entries.
It is this CS matrix yielding the Jain hierarchy.

The   IIB picture 
for Jain hierarchy looks as follows. Consider
p D3 branes stretched between (0,1) and (1,1) 5 branes.
There is also nonzero density of D1 branes 
on D3 branes which make theory noncommutative.
This configuration corresponds to 
the integer QHE effect 
with filled p Landau levels. Consider now m
additional semiinfinite D3 branes on the right
from (1,1) 5 brane and  far from the
stretched D3 branes along the $(x_4,x_5)$ coordinates.
As usual, the distance along  $(x_4,x_5)$ coordinates
between stretched and semiinfinite D3 branes
corresponds to the masses of the latter.
Therefore the  semiinfinite branes 
correspond to the adding of the
matter in the fundamental with the very large mass.
Note that the brane configuration implies that  
auxiliary fermions  fill m Landau levels.
If we integrate out semiinfinite D3 branes
the CS level gets renormalized and 
we just arrive at the effective Jain picture (see Fig. 1 a)
In principle the auxiliary fundamentals
could be also represented by  m D5 branes localized in $x_6$
instead of the semiinfinite D3 branes. However
the latter picture  is more suggestive
for a further generalization to more generic 
hierarchies. 
\begin{figure}
\begin{center}
\epsfxsize=5in
\epsfbox{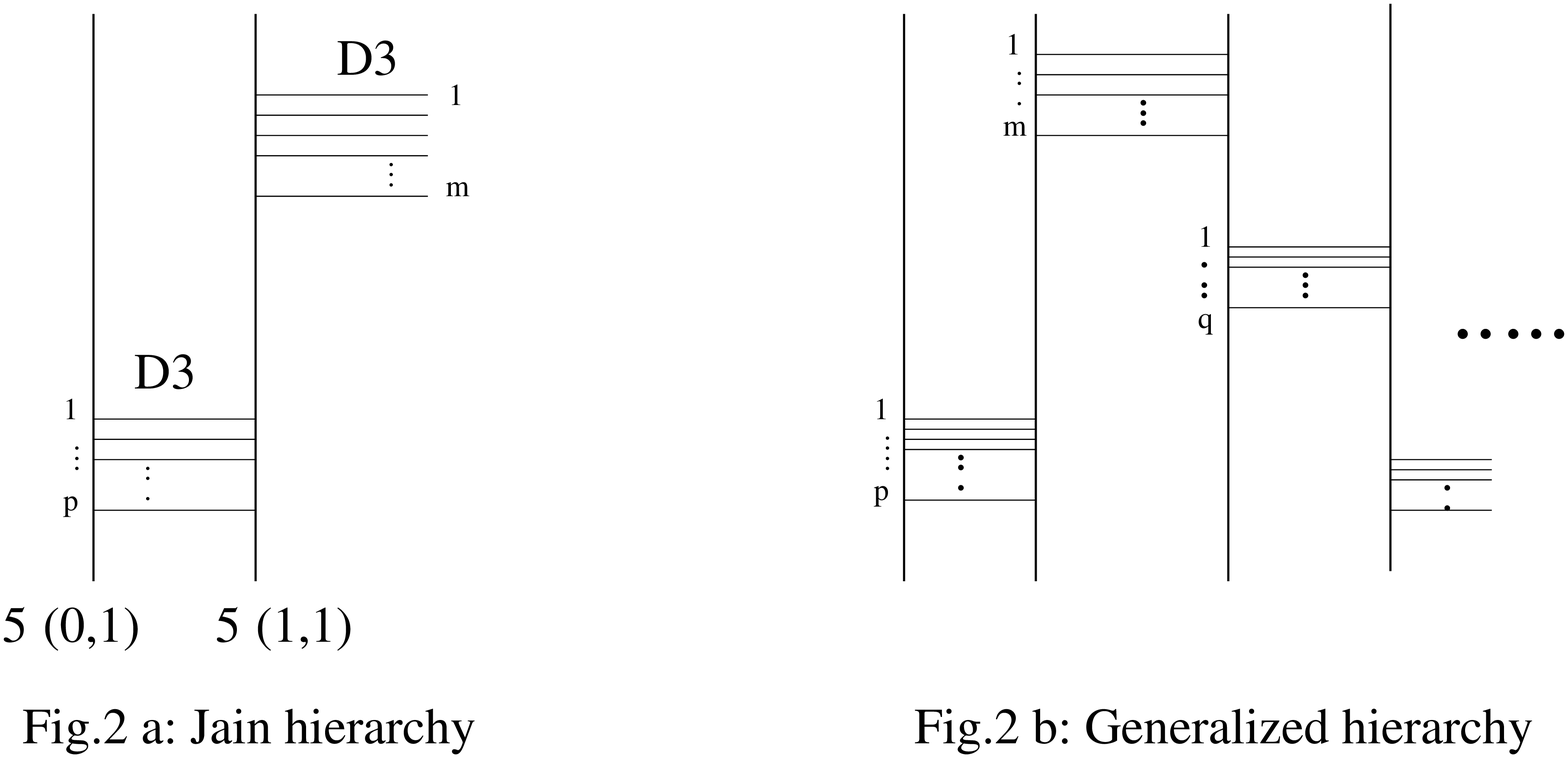}
\caption{Brane configuration for Jain (Fig.2a) and generalized (Fig.2b) 
 hierarchies}
\end{center}
\end{figure}

\subsection{Quiver brane models and the generalized hierarchies} 

The brane configuration suggested for an Jain 
hierarchy implies the immediate generalization.
Let us consider the IIB brane configuration
relevant for the generalized hierarchy. Our starting
point is the quiver type brane  configuration which
involves some number of 
$(p_i,q_i)$ 5 branes 
located at points $x_{6i}$ and $m_i$ D3 branes
stretched between i-th and (i+1)-th 5 branes.
There are also D1 branes inside D3 branes 
representing the electrons. Therefore 
the gauge theory on the worldvolume of D3 branes
has the gauge group $\prod SU(m_i)$ with
bifundamental matter which corresponds to the
fundamental string with ends on different stacks
of D3 branes  
(see Fig. 1b). Once again the nontrivial density
of D1's on D3 branes is implied.
This  configuration is the generalization of our picture
for the Jain hierarchy. The levels of CS terms
induced in $SU(m_i)$ factors   
are 
\beq
k_i= \frac{p_i q_{i+1} -p_{i+1} q_{i}}{q_iq_{i+1}}
\eeq 

Let us suggest the following 
generalized mechanism of the condensation
similar to the discussed above for the Jain case.
The matrix $K$ is given by the same formula as before 
\beq
K_{ij}= K^0_{ij} + 2\nu_1 (1)_{ij}
\eeq
where the matrix  $K^0_{ij}$ is defined 
by the charges of the first and the second 5 branes. 
However the 
renormalization 
of the Chern-Simons level is much more involved. 
At the first step instead of integer filling factor $m$  which we had  
from  $m$ semiinfinite $D3$  branes we now have the filling factor
for $m$ finite $D3$ branes stretched between second and third five-branes.
This filling 
factor is determined by its matrix $K^{(1)}$ 
which has the contribution  $K^{(1),0}_{ij}$     amounted from
the charges of the second and third 5 branes  
and the next level renormalization
\beq
K^{(1)}_{ij}= K^{(1),0}_{ij} + 2\nu_2 (1)_{ij}
\eeq
At the next step the filling factor 
for another stack of $D3$ branes between third 
and fourth five branes will be defined through it own matrix $K^{(3)}$ which 
in turn is defined through  the next one, etc, etc
\beq
K^{(n)}_{ij}= K^{(n),0}_{ij} + 2\nu_{n+1} (1)_{ij}
\eeq
Therefore the whole filling factor is defined 
by the set of integers $(p_i,q_i,m_i)$. To have
some step by step renormalization some ordering
of the bifundamental masses should be imposed.

As an example, consider the simple case when
all  $K^{(n),0}_{ij}=\delta_{ij}$. This 
can be easily achieved assuming that only 
(1,0) and (1,1) 5 branes are involved.
One can see that this will lead to the chain fraction
\beq
\nu = 1/\left( 1/p + 2/\left( 1/m + 2\left(1/q +.... \right)\right)\right)
\eeq
which can be terminated at some finite level N when we send all 5-branes 
starting from brane number $N+1$ to infinity.

\section{Duality in FQHE and IIB brane picture}
\subsection{Mirror transformation}

Is Section 4.1 we have recalled that there are three 
natural transformations relevant for the duality
group in FQHE. 
Hence let us discuss now how these transformations
could be recognized in IIB picture. First,
consider the motion of 5 branes which has the Seiberg
duality between two N=1 SUSY gauge theories in the IR as
the four dimensional counterpart. We will argue that 
this motion corresponds to the particle-hole transformation
of the FQHE. Let us start with the simplest situation
with two two $(p_1.q_1)$ and $(p_2,q_2)$ 5 branes located
differently along $x_6$ coordinate  and N D3 branes 
stretched between them. 

Consider the interchange of
the 5 branes. It was argued in \cite{csbranes} that
after this transformation the number of D3 branes 
becomes $|p_1q_2 - p_2q_1|- N$. Due to the brane 
motion the effective filling fraction of the
corresponding Quantum Hall systems changes.
Indeed, assume for the simplicity that $q_1=q_2=1$,
then before the transition the effective 
filling factor in $\nu= \frac{N}{p_1 - p_2}$. After
the transition the effective level 
of the CS term does not change, however
the rank of the gauge group does. Therefore
the new effective filling factor is 
$\tilde{\nu}=1 - \nu$ that is this
mirror transform corresponds to the
particle-hole transformation.
 
Let us argue that a similar answer can be
derived in the response model for the
Jain hierarchy. Assume 
that we are dealing with (1,0) and (1,1)
5 branes providing the level k=1 for
each $U(1)^k$ factor (k=1,..,N). There are
also m semiinfinite D3 branes or D5 branes 
representing the auxiliary fermions. When 
5 branes are moved through each other 
just as in Seiberg duality in N=1 D=4
SUSY Yang-Mills theory 
the dual gauge group SU(m-N) emerges 
which results in the same transformation
of the filling factor.

\subsection{S duality and the transitions between plateaux }

One more duality natural in the brane setup
is the stringy S duality transformation.
In IIB picture S duality is the part 
of the known SL(2,Z) duality group. 
As we have already have mentioned
the natural modular parameter in the Quantum Hall context 
is the complex conductivity $\sigma$.
In the IIB brane picture it is natural 
to assume that 
$\sigma_{xx} \propto 1/g^2$ 
and therefore is proportional to the
distance between 5 branes along $x_6$ coordinate.
The picture is actually similar to the 
IIA brane presentation of the N=1 SUSY theories.
Let us  recognize the S duality transformations
in the brane setup.

Let us
start first with the abelian CS theory 
with m auxiliary fundamentals which
correspond to the Laughlin states.
S duality maps NS5 brane into D5 brane
while (p,q) 5 brane is mapped into (q,-p)
5 brane. 
If there are no auxiliary fermions
(m=0) then S duality amounts to the
transformation from the QHE duality group
$\nu \rightarrow \frac{1}{\nu}$.
When $m\neq 0$ the situation is slightly
more involved. The corresponding gauge group
appears to be $U(1)^{m-1}\times U(1)$
where the CS level is nonzero only for the
last gauge factor.
Therefore the Jain hierarchy gets mapped
into the generalized hierarchy under 
S-duality.

The next question concerns taking into
account
the gauge coupling constant which 
we have to  consider in the full  Chern-Simons-Maxwell
lagrangian. It was argued in \cite{csbranes}
that the natural RS transformation which involves
S duality as well as the flip with respect to
the three coordinates  (789) maps the theory with
the CS level k into the dual theory
with the level $\tilde {k}= - \frac{1}{k}$
while the coupling constant is mapped into 
the mass of the dual gauge boson. The partition
functions of these theories are proportional
to each other therefore we see that the account
of the gauge coupling proportional to $\sigma_{xx}$
does not destroy the duality group. Let us note that
the same transformation $\tilde {k}= - \frac{1}{k}$
can be obtained in just $2+1$ topologically massive 
gauge theory also as a mirror transformation and it can be shown
how  this $2+1$ CS mirror transform is related to $T$ and $S$-dualities
in string theory \cite{CSmirror}

It is worth to determine clearly the meaning of
two remaining basic transformations from 
the duality group $\Gamma_{U}(2)$, namely
the Landau level attachment(L) and the flux 
attachment(F). In our IIB picture their
interpretations are evident;the  L transformation
just adds(or removes) one stretched D3 brane 
to the configuration while the F transformation
adds (remove) the semiinfinite D3 brane. 
Let us emphasize that the consistency
conditions of such transformations with the RG flows 
known in FQHE can  be expected. Indeed, it is a kind of 
the consistency condition for the 
decoupling of the heavy matter standard in the
field theory context. When one tries to decouple
the heavy matter the dynamically generated scales
in the theory have to be changed due to the 
change of the $\beta$ function in the theory.

Another issue which could be questioned in terms
of the IIB brane picture is the transition between
plateaux. The corresponding RG flows certainly 
involve the gauge coupling constant in the 
gauge theory description. We  conjecture
that the transition can be described in terms
of a web of (p,q) 5 branes. Indeed, there
exist  web like configurations when the
5 branes change their quantum
numbers passing through 
the junction 
manifolds. In terms of the gauge theory on D3
branes this means the change of levels 
of CS terms and therefore of filling
fractions. After the junction manifold
the emerging D3 brane is tilted between
other 5 branes. The process of the motion
of D3 brane through the 5 brane web
on the other hand can be interpreted as the 
nontrivial renormalization of the coupling
constant.
\begin{figure}
\begin{center}
\epsfxsize=5in
\epsfbox{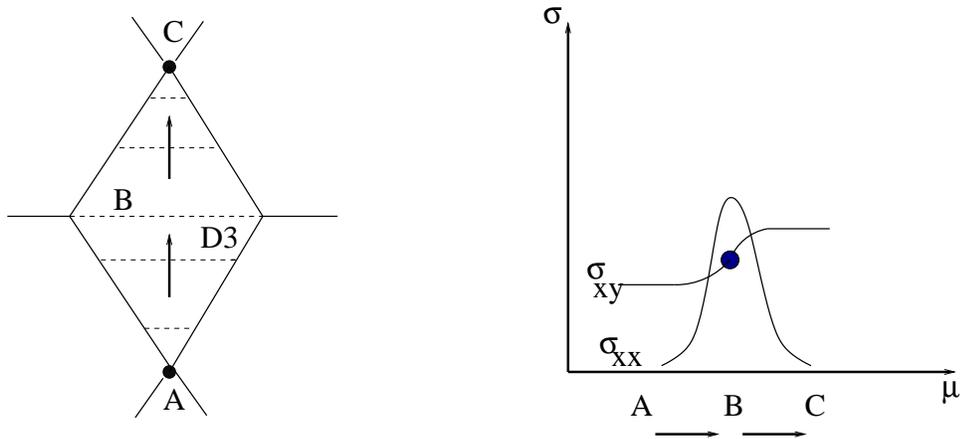}
\caption{Brane  realization of the transition between plateaux. 
In the position B one  gets a conformal field theory}
\end{center}
\end{figure}

Let us explain the simplest example 
of such transition a little bit more
precise. Suppose that  5 branes
are originating in a "diamond" configuration
(see Fig.3). 
\begin{figure}
\begin{center}
\epsfxsize=5in
\epsfbox{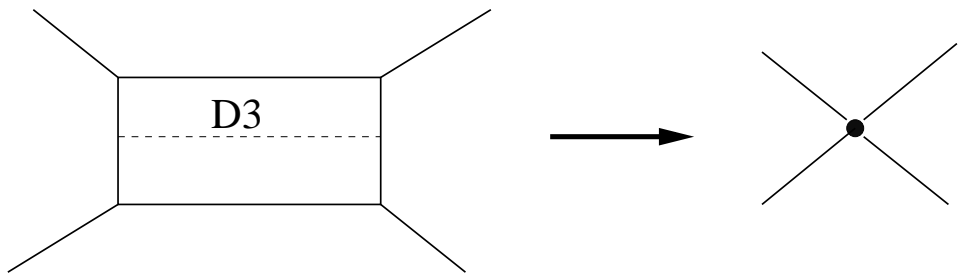}
\caption{Box made from  five-branes with a D3 brane inside.}
\end{center}
\end{figure}

The crucial point is that 
is was argued that any intersection
of (p,q) 5 branes amounts to the nontrivial
superconformal point in d=5 theory on
5 branes \cite{ah}. Moreover if the D3 brane
is stretched between 5 branes when D3 brane 
moves to the these intersection points the
critical d=3 theories emerge as well. Let
us try to exploit this setup to explain 
the transitions. Our conjecture is formulated as 
follows. Since the quantum numbers 
of 5 branes fix the filling factors
uniquely we start with position of D3 brane
at the down vertex of the diamond. This 
vertex correspond to the initial 
fixed point of the QH RG flow at some plateau.
It is clear from the conjectured brane geometry that
$\sigma_{xx}=0$ at this point in agreement
with the QH flows. More accurately 
to see the nontrivially generated CS level
due to the five brane quantum numbers
it is necessary to assume 
that the vertex itself has the 
internal microscopic structure (see Fig.4)

Then 
we move D3 brane up interpreting this motion
as the RG flow. At some value of $x_3$ D3 
brane approaches the next critical point
corresponding the point of two 5 brane junctions.
This critical point geometrically corresponds to    
$\sigma_{xx} \neq 0$ . 
This is consistent with the existence 
of the conformal fixed point at $\sigma_{xx} \neq 0$
between the plateaux.
Recently it was argued that 
this  conformal point 
corresponds to the noncompact 
SL(2,R) WZW model \cite{zirn,kogan2}. Since the group
SL(2,R) geometrically is $AdS_{3}$ we
would like to ask where this $AdS_{3}$
could come from. Fortunately it can be easily
found. Indeed there is nonzero density of D1
branes on D3 branes. On the other hand it is
known that N D1 branes at large N creates 
$AdS_{3}$ geometry needed 
for the WZW model. This coincidence 
can be considered as a rough consistency
check of our conjecture.
Finally D3 brane 
reaches the upper vertex of the diamond 
with the other filling factor which
can be calculated using the conservation
of the 5 brane charges at the junction manifold.

One more evident check follows from the known
logarithmic renormalization of  $\sigma_{xx}$
which has to be seen 
in the brane picture geometrically. Namely 
in analogy with coupling constant renormalization
in N=1 SYM theory we have to interprete
this running coupling constant as the solution to 
the Laplace equation with a source, represented
by D3 brane in $x_4,x_5$ coordinates. Removing
one D3 brane from the vertex we destroy the
criticality and it is natural to assume 
that  $\beta$ function comes precisely from 
this single D3 brane. If this interpretation
is true the independence of the weak coupling 
$\beta$ function on the filling fractions 
asquires the natural explanation.

Note that the argumentation above was applied 
to the transitions between the integer Quantum Hall
plateaux which are represented by the D3 branes
with the finite extent in $x_6$ direction. The 
simplest transition between these plateuax corresponds 
to removing one D3 brane with some number 
of D1 branes from the bunch of the remaining branes. 
Hence, we have
indeed the motion of D1 brane in the background 
of many D1's providing the $AdS_3$ geometry. To
consider the generic transitions between FQHE 
plateaux we have to take into account the 
motion the semiinfinite D3 branes along the 
five-brane web. It is still unclear what universality
class governs the generic transitions between FQHE
plateaux.

\section{Discussion}
In this paper we discuss several aspects of the
realization of the FQHE via noncommutative Chern-Simons
theory. First, we generalize the description 
of the FQHE system with the finite amount of electrons
to the topology of the torus. We argued that the 
corresponding many-body system can be identified with
the integrable Ruijsenaars model. Then we interpret
two alternative descriptions of the model in terms
of commutative and noncommutative gauge theories
as the version of Morita duality.

Using the brane description of the many-body system
we suggest a new IIB brane picture for the FQHE system
which is based on the brane realization of CS terms
via (p,q) 5 branes. It is quite transparent and
can be easily generalized for the Jain and Haldane
hierarchies. Moreover it implies some dualities
which partially have counterparts in the FQHE system
with finite or infinite number of electrons. 
In particular S duality as well as mirror symmetry
can be identified.
A brane realization 
of the RG flows was conjectured.

Finally one could ask even more general 
questions concerning the physical meaning
of the extra coordinates in the context
of the QHE. In  SUSY gauge theories
these coordinates come from the higher
dimensional components of the gauge
field which are interpreted as the 
scalar fields on the brane worldvolume.
Naively there is no place for 
the additional scalar fields in the
context of QHE so one should look
for another viewpoint. 

We can speculate 
that the additional coordinates actually
could be related with the momentum space
of the usual four-dimensional theory. 
Then the corresponding brane picture
can be interpreted in terms of the
Fermi surfaces similar to the 
interpetation of the Peierls model
in \cite{pei}. The set of branes in the
momentum space can be treated along this way
as the result of the complicated level
crossing phenomena which are known to
provide the defects of the different codimensions
resulting in  nontrivial Berry phases.  The size of the brane is
related to the chemical potential. Let us note that in the ``diamond''
picture on Fig.3 the coordinate describing the motion in the vertical 
direction  is related to the chemical potential. Brane motion in this
direction corresponds to the transition between plateaux which in turn
leads to the change in chemical potential. Note that with
this interpretation we have no reference whatsoever 
to the Planck scale.

Let us emphasize once again that there
are similarities between the brane 
descriptions of QHE and N=1 SUSY theories.
Both theories manifest the behaviour with
the dynamically generated mass gap. In
$D=4$ N=1 this mass scale can be seen in
 the MQCD description geometrically. The same 
situation we expect in the  M theory description
of FQHE however 
to see this more explicitly 
the additional analysis is required. 
And most importantly both theories
being the low energy ones have nothing to
do with the quantum gravity scale and all degrees of
 quantum gravity scale are decoupled. This is absolutely
 important for the consisteny of the whole brane approach to
 the physical theories with a  well-defined UV behaviour -
 obviously QHE as well as QCD are exactly such theories.  
 
 Finally let us briefly discuss possible  brane 
 description of  CS  of  (quasi)planar superconducting state. 
 It was suggested  some time ago that anyon system can be in 
 a superconducting state\cite{superconductivity}. 
 The transition to this state
 is due to emergence of massless mode - a $3D$ photon. 
 The mass of the photon in a topologically massive
 gauge (Maxwell-CS) theory is proportional to CS coefficient $k$ and in
 superconducting  state bare $k$ is completely cancelled  by one-loop
 correction (actually it
 happens only at zero temperature, at nonzero $T$ there is still small
 mass \cite{nathan}). It will be interesting to have brane realiztion 
for  anyon  superconductivity  and to find what integrable model it 
corresponds to 
\footnote{
When this paper was prepared for publication we
 became aware of a recent paper \cite{asorey} where CS theory was used
 to describe BCS system. In the case of QHE the CS description was given
 by $SU(N)$ (where $N$ is the number of electrons) with one Wilson line
 where in BCS case it is  $SU(2)$ with M Wilson lines (where $M$ is the 
 number of the Cooper pairs ). It is amusing that there is a connection
 between $SU(N)$ gauge theory on the surface 
with one Wilson line and $SU(2)$ gauge theory  with N Wilson
 lines. This relation 
has the interpretation in terms of the
separation of  variables procedure 
in the integrability context (see \cite{gnr} and references therein).
}.
 In this brane picture
 we must have an effect of complete (at $T =0$) or partial (at $T \neq
 0$) screeing of CS coupling.
 It will be also interesting to find brane realization
 of more complicated superconduction systems, like P-even anyon 
superconductors \cite{kovner}.

The work of A.G. is supported in part
by grants INTAS-00-00334  and CRDF-RP1-2108,
IK is supported in part by PPARC 
rolling grant PPA/G/O/1998/00567,
both I.K. and C.K.A. are supported in part by the EC TMR grant
HPRN-CT-2000-00152 and a joint  CNRS-Royal Society grant.

\end{document}